\documentclass[12pt]{iopart}
\usepackage{graphicx}
\usepackage{iopams}  
\begin{document}

\title{Motion of charged particles in an electromagnetic knot}

\author{M. Array\'as and J. L. Trueba}

\address{\'Area de Electromagnetismo, Universidad Rey Juan
Carlos, Camino del Molino s/n, 28943 Fuenlabrada, Madrid, Spain}

\begin{abstract}
In this paper we consider the classical relativistic motion of charged particles 
in a knotted electromagnetic field. After reviewing how to construct electromagnetic knots
from maps between the three-sphere and the two-sphere, we introduce a mean quadratic 
radius of the energy density distribution in order to study some properties of this field.
We study the classical relativistic motion of electrons in the electromagnetic field of the Hopf map,
and compute their trajectories. It is observed that these electrons initially at rest are strongly accelerated by 
the electromagnetic force, becoming ultrarelativistic in a period of time that depends on the knot 
energy and size.
\end{abstract}

\vspace{2pc}
\submitto{\JPA}
\maketitle

\section{Introduction}

As pointed out in a recent paper by Irvine and Bouwmeester
\cite{Irv08}, electromagnetic knots are exact solutions of the
classical Maxwell equations of electromagnetism in vacuum. They
appeared firstly in a paper by Ra\~nada \cite{Ran89} in 1989.
Ra\~nada himself \cite{Ran92} has used these solutions as the basic elements of a
topological model of electromagnetism which is locally equivalent to Maxwell's standard
theory but implies furthermore some topological quantization
conditions with interesting physical meaning
\cite{Ran95,Ran97,Ran98,Ran03,Ran06}.

Electromagnetic knots are defined through two fundamental complex scalar fields
$(\phi ,\theta )$ whose level curves coincide with the
magnetic and electric lines respectively, each one of these lines
being labelled by the constant value of the corresponding scalar.
Both scalars are assumed to have only one value at infinity,
which is equivalent to compactify the physical three-space to the
sphere $S^3$. Moreover, the complex plane is compactified
to the sphere $S^2$ via stereographic projection. As a result
of such compactifications, the scalars $\phi$ and $\theta$ can be 
interpreted, at any time, as maps $S^3\rightarrow
S^2$, which can be classified in homotopy classes, characterized by 
the value of the Hopf index $n$ \cite{Hopf}. It can be shown
that the two scalars have the same Hopf index and that the
magnetic and the electric lines are generically linked with the
same Gauss linking number $\ell$. If $\mu$ is the
multiplicity of the level curves (i.e. the number of different
magnetic or electric lines that have the same label $\phi$ or
$\theta$), then the Hopf index of both scalars is $n=\ell \mu ^2$. 
The Hopf index can thus be interpreted as a generalized linking number 
if we define a line
as a level curve with $\mu$ disjoint components.

An important feature of the model is that the Faraday 2-form
${\cal F} = \frac{1}{2} F_{\mu \nu}dx^\mu \wedge dx^\nu$ and its
dual $*{\cal F}=\frac{1}{2}\, ^*\!F_{\mu \nu}dx^\mu \wedge dx^\nu$
are proportional to the two pull-backs of $\sigma$, the area 2-form in $S^2$, by
$\phi$ and $\theta$,
\begin{equation}
{\cal F} = - \sqrt{a} \, \phi ^*\sigma , \, \, \, *{\cal F} = c \sqrt{a} \, \theta ^*\sigma ,
\label{pullback}
\end{equation}
where $a$ is a constant introduced so that the magnetic and electric fields have correct
dimensions and $c$ is the velocity of light in vacuum. In the International System of Units, 
$a$ can be expressed as a pure number times the Planck constant $\hbar$ times the light 
velocity $c$ times the vacuum permeability $\mu_{0}$.
As a consequence of the definitions (\ref{pullback}), the maps $\phi$ and 
$\theta$ are dual to one another, $*(\phi ^*\sigma )=-\theta ^*\sigma$, where
* is the Hodge or duality operator. This duality condition
guarantees that both $\cal F$ and $*\cal F$ obey the Maxwell
equations in empty space without the need of any other
requirement. 

The electromagnetic fields obtained in this way are
called electromagnetic knots. They are radiation fields as
they verify the condition ${\bf E}\cdot {\bf B}=0$. It can be 
proved (see \cite{Ran97} for the details) that any radiation field 
in vacuum is locally equivalent to an electromagnetic knot. 
Moreover, because of the Darboux theorem, any electromagnetic
field in empty space can be expressed locally as the sum of two
radiation fields. Consequently, a model of electromagnetism based on these
electromagnetic knots is locally equivalent to Maxwell standard
theory. However, its difference from the global point of view has
interesting consequences, as are the following topological
quantizations: (i) The electric charge of any point particle must
necessarily be equal to an integer multiple of the
fundamental value $q_0=\sqrt{\hbar c \epsilon _0}$ (see \cite{Ran98}). 
(ii) The
electromagnetic helicity ${\cal H} = \hbar c (N_{R} - N_{L})$ is
also quantized \cite{Ran97}, where $N_R$ and $N_L$ are the
classical expressions of the number of right- and left-handed
photons contained in the field (i.e. $N_R-N_L=\int d^3k (\bar{a}_R
a_R-\bar{a}_L a_L)$, $a_R({\bf k}),a_L({\bf k})$ being Fourier
transforms of the vector potential $A_\mu$ in the classical
theory, but creation and annihilation operator in the quantum
version). In fact, for any electromagnetic knot, $n= N_R-N_L$,
which is a remarkable relation between the Hopf index (i.e. the
generalized linking number) of the classical field and the
classical case of the difference $N_R-N_L$. (iii) The topology 
of the model implies
also the quantization of the energy of the electromagnetic field
in a cavity \cite{Ran03}. (iv) Electromagnetic knots
are compatible with the quantization of the magnetic flux of a
superconducting ring \cite{Ran06}, which in standard theory is
always an integer multiple of $g/2$, $g$ being the value of the
magnetic monopole.

In this work we study a classical charged particle in a knotted
electromagnetic field. We find that the particles can accelerate to light velocity. 
We first revise the construction and some physical propeties of the electromagnetic
field built from the Hopf map between the compactified three-space (the sphere $S^3$) 
and the compactified complex plane (the sphere $S^2$). In particular, we pay attention 
to the electromagnetic energy density and how it evolves with time and introduce
a mean quadratic radius of the energy density. Then we consider
the relativistic motion of electrons in this
electromagnetic field. We study the trajectories of the electrons and their velocities, 
showing that they become ultrarelativistic for a wide range of the electromagnetic energy
of the knot. Finally we give some conclusions and prospects of future work.

\section{The Electromagnetic field of the Hopf fibration}

A method to find explicitly some electromagnetic knots can be
found in \cite{Ran01}. Let $\phi_{0} ({\bf r})$, $\theta_{0} ({\bf
r})$, two complex scalar fields such that can be considered as
maps $\phi_{0}, \theta_{0} : S^3 \rightarrow S^2$ after identifying the
physical space $R^3$ with $S^3$ and the complex plane with $S^2$.
They have to satisfy the following two conditions:

1. The level curves of $\phi_{0}$ must be orthogonal, in each
point, to the level curves of $\theta_{0}$, since we know that
electromagnetic knots are radiation fields (${\bf E}\cdot {\bf
B}=0$).

2. The Hopf index of $\phi_{0}$ and of $\theta_{0}$ are equal,
$H(\phi_{0})=H(\theta_{0})$. This is necessary to ensure that the
condition ${\bf E}\cdot {\bf B}=0$ is maintained during time
evolution.

Given $\phi_{0}$ and $\theta_{0}$ with this two conditions, we can build
the magnetic and electric fields at $t=0$ as
\begin{eqnarray} 
{\bf B}({\bf r},0) &=&\frac{\sqrt{a}}{2\pi
i}\frac{ \nabla \phi_{0} \times \nabla {\bar{\phi}}_{0}}{(1+
{\bar{\phi}_{0}}\phi_{0} )^{2}},  \nonumber \\
{\bf E}({\bf r},0) &=&\frac{\sqrt{a} c}{2\pi
i}\frac{\nabla {\bar{\theta}}_{0}\times
\nabla \theta_{0}}{(1+{\bar{\theta}}_{0}\theta
_{0})^{2}}. \label{11.3}
\end{eqnarray}
It is convenient to work with dimensionless coordinates in the
mathematical spacetime $S^3 \times R$, and in $S^2$. In order to
do that, we define the dimensionless coordinates $(X, Y, Z, T)$,
related to the physical ones $(x, y, z, t)$ (in the SI of units that we will use
in this work) by
\begin{equation}
(X, Y, Z, T) = \frac{1}{L_{0}} (x, y, z, c t) ,  \label{11.18}
\end{equation}
and $r^2 /L_{0}^2 =(x^2 + y^2 +z^2)/L_{0}^2= X^2 + Y^2 +Z^2
=R^2$, where $L_{0}$ is a constant with length
dimensions. Now, let us consider the Hopf map,
\begin{equation}
\phi _{0}=\frac{2(X+iY)}{2Z+i(R^{2}-1)},  \label{11.19}
\end{equation}
whose fibres are been used as a basis for a case of knotted entanglement in reaction-diffusion 
models, in particular for a FitzHugh-Nagumo model \cite{Tru09}. We also consider the map 
corresponding to the change $(X,Y,Z)\mapsto (Y,Z,X)$
in (\ref {11.19}),
\begin{equation}
\theta _{0}=\frac{2(Y+iZ)}{2X+i(R^{2}-1)}.  \label{11.20}
\end{equation}
Because of their construction, it is obvious that both maps
(\ref{11.19}) and (\ref{11.20}) have the same Hopf index. In fact,
these maps have Hopf index $n=1$ and their fibrations are mutually
orthogonal at each point. Consequently, we can build an
electromagnetic knot from these maps. The Cauchy data for the
magnetic and electric fields are
\begin{eqnarray}
{\bf B}({\bf r},0) &=&\frac{8\sqrt{a}}{\pi L_{0}^2
(1+R^{2})^{3}}\left(
Y-XZ,-X-YZ,\frac{-1-Z^{2}+X^{2}+Y^{2}}{2}\right)  , \nonumber \\
{\bf E}({\bf r},0) &=&\frac{8\sqrt{a} c}{\pi L_{0}^2
(1+R^{2})^{3}}\left( \frac{1+X^{2}-Y^{2}-Z^{2}}{2},-Z+XY,Y+XZ \right) .
\label{11.27}
\end{eqnarray}
From (\ref{11.27}), two vector potentials ${\bf A}$ and ${\bf C}$
can be computed, such that ${\bf B}={\mbox{\boldmath$\nabla$}}
\times {\bf A}$, ${\bf E}={\mbox{\boldmath$\nabla$}} \times {\bf
C}$, with the results
\begin{eqnarray}
{\bf A}({\bf r},0) &=&\frac{2\sqrt{a}}{\pi L_{0}
(1+R^{2})^{2}}\left(
Y,-X,-1\right) ,  \nonumber \\
{\bf C}({\bf r},0) &=&\frac{2\sqrt{a} c}{\pi L_{0}
(1+R^{2})^{2}}\left( 1,-Z,Y\right)  . \label{11.28}
\end{eqnarray}
The magnetic and electric helicities of this knot are defined to be
\begin{eqnarray}
h_{m}=\frac{1}{2 \mu_{0}} \int_{R^{3}}{\bf A}\cdot {\bf B}\,d^{3}r, \nonumber \\ h_{e}=
\frac{\varepsilon_{0}}{2} \int_{R^{3}}{\bf C}\cdot {\bf
E}\,d^{3}r, \label{11.29}
\end{eqnarray}
where $\varepsilon_{0}$ is the vacuum permittivity. Taking into account the Cauchy data (\ref{11.27}) 
and the potentials (\ref{11.28}), the electromagnetic helicity yields
\begin{equation}
h = h_{m} + h_{e} =  \frac{a}{2 \mu_{0}} + \frac{a}{2 \mu_{0}} = \frac{a}{\mu_{0}}. \label{11.299}
\end{equation}
To find the electromagnetic knot at any time from the
Cauchy data (\ref{11.27}), we use Fourier analysis. The
fields turn out to be \cite{Ran01},
\begin{eqnarray}
{\bf B}({\bf r},t) &=&\frac{\sqrt{a}}{\pi L_{0}^2(A^{2}+T^{2})^{3}%
}\left( Q{\bf H}_{1}+P{\bf H}_{2}\right) ,  \nonumber \\
{\bf E}({\bf r},t) &=&\frac{\sqrt{a} c}{\pi L_{0}^2
(A^{2}+T^{2})^{3}}\left( Q{\bf H}_{2}-P{\bf H}_{1}\right) ,
\label{11.35}
\end{eqnarray}
where the quantities $A$, $P$, $Q$ are defined by
\begin{equation}
A=\frac{R^{2}-T^{2}+1}{2},P=T(T^{2}-3A^{2}),Q=A(A^{2}-3T^{2}),
\label{11.36}
\end{equation}
and the vectors ${\bf H}_{1}$ and ${\bf H}_{2}$ are
\begin{eqnarray}
{\bf H}_{1} &=&\left( Y+T-XZ,-X-(Y+T)Z,\frac{-1-Z^{2}+X^{2}+(Y+T)^{2}}{2}%
\right)  , \nonumber \\
{\bf H}_{2} &=&\left( \frac{1+X^{2}+Z^{2}-(Y+T)^{2}}{2},-Z+X(Y+T),Y+T+XZ%
\right)  . \label{11.37}
\end{eqnarray}
This solution fulfills now ${\bf E} \cdot {\bf B} =0$ and
$E^{2}- c^{2} B^{2}=0$ at any time. It is possible to obtain directly the electromagnetic field
(\ref{11.35}) from the time-dependent expressions (\ref{pullback}). In terms of the 
magnetic and the electric fields, we have
\begin{eqnarray}
{\bf B}({\bf r},t) &=&\frac{\sqrt{a}}{2 \pi i (1 + \phi {\bar \phi} )^2} 
\nabla \phi \times \nabla {\bar \phi} = \frac{\sqrt{a}}{2 \pi i c (1 + 
\theta {\bar \theta} )^2} \left( \frac{\partial {\bar \theta}}{\partial t} 
\nabla \theta - \frac{\partial \theta}{\partial t} 
\nabla {\bar \theta} \right) , \nonumber \\
{\bf E}({\bf r},t) &=&\frac{\sqrt{a} c}{2 \pi i (1 + \theta {\bar \theta} )^2} 
\nabla {\bar \theta} \times \nabla \theta = \frac{\sqrt{a}}{2 \pi i (1 + 
\phi {\bar \phi} )^2} \left( \frac{\partial {\bar \phi}}{\partial t} 
\nabla \phi - \frac{\partial \phi}{\partial t} 
\nabla {\bar \phi} \right) ,
\label{timeeb}
\end{eqnarray}
where the time-dependent expressions of the maps $\phi$ and $\theta$ are (see \cite{Ran97})
\begin{eqnarray}
\phi &=& \frac{(AX-TZ)+i(AY+T(A-1))}{(AZ+TX)+i(A(A-1)-TY)}, \nonumber \\
\theta &=& \frac{(AY+T(A-1))+i(AZ+TX)}{(AX-TZ)+i(A(A-1)-TY)}.
\label{timemap}
\end{eqnarray}
Note that the level curves of both maps $\phi$ and $\theta$ remain linked with a 
Gauss linking number equal to 1. The evolution of the curved lines of these maps
gives the evolution of the force lines of the magnetic and electric fields given
by the expressions (\ref{timeeb}).

\begin{figure}
\centering
\includegraphics[width=0.45\textwidth]{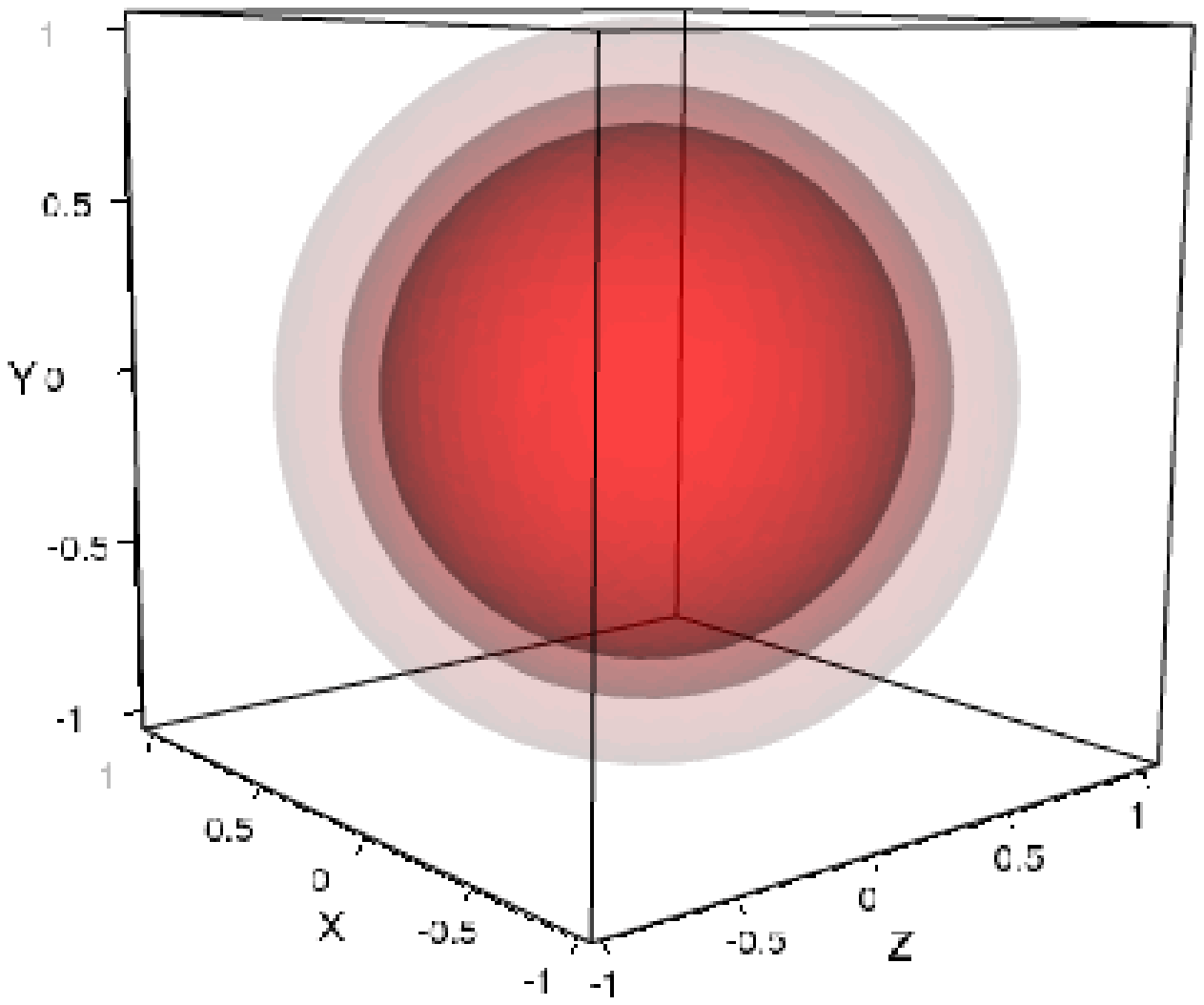}
\includegraphics[width=0.45\textwidth]{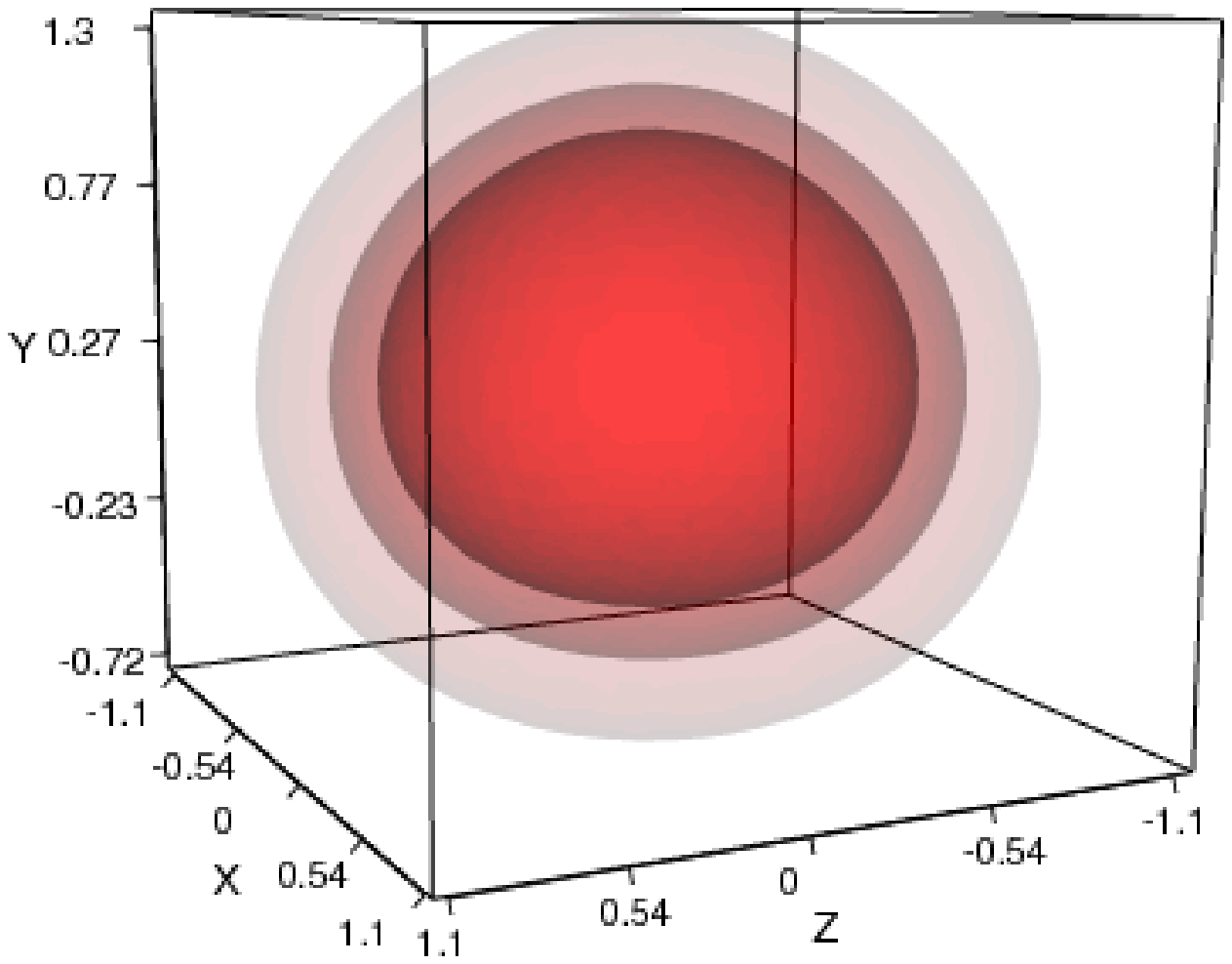}
\includegraphics[width=0.45\textwidth]{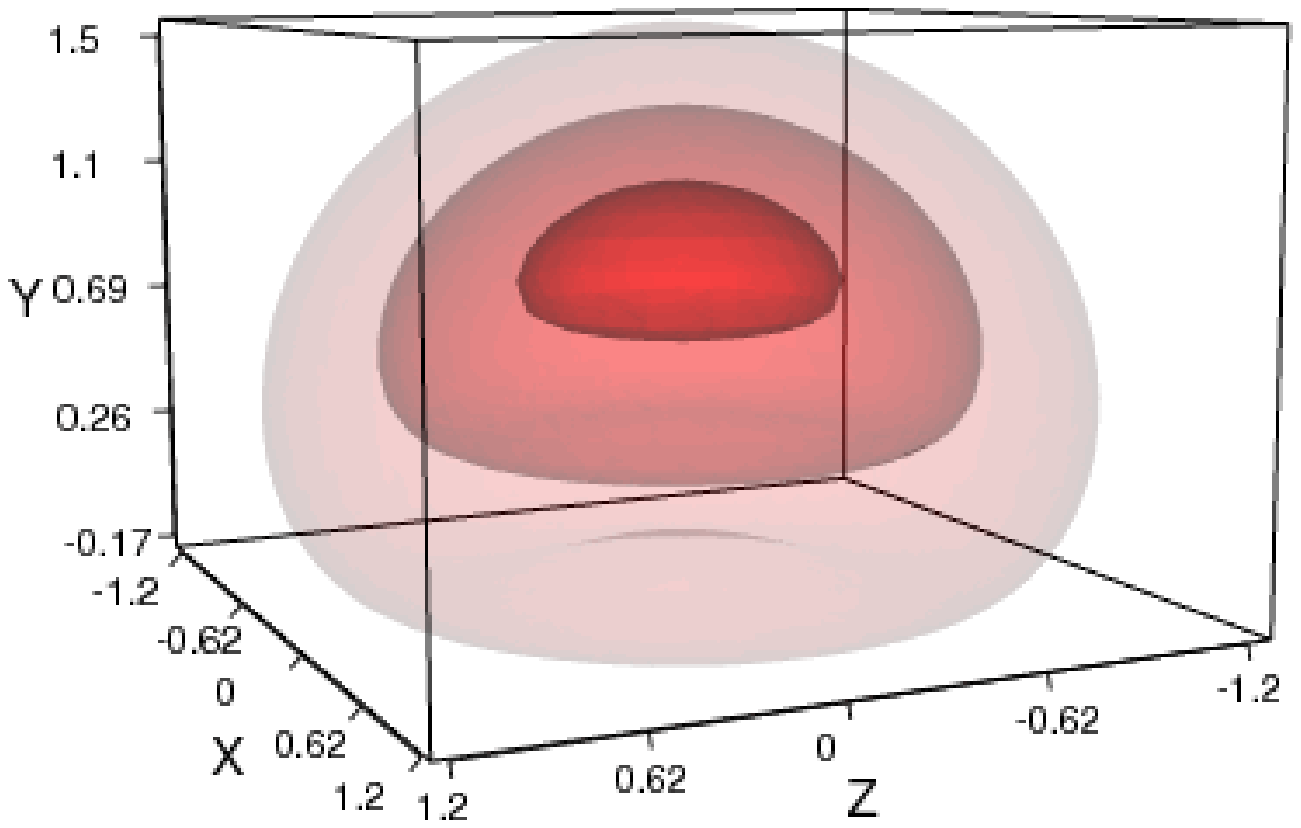}
\includegraphics[width=0.45\textwidth]{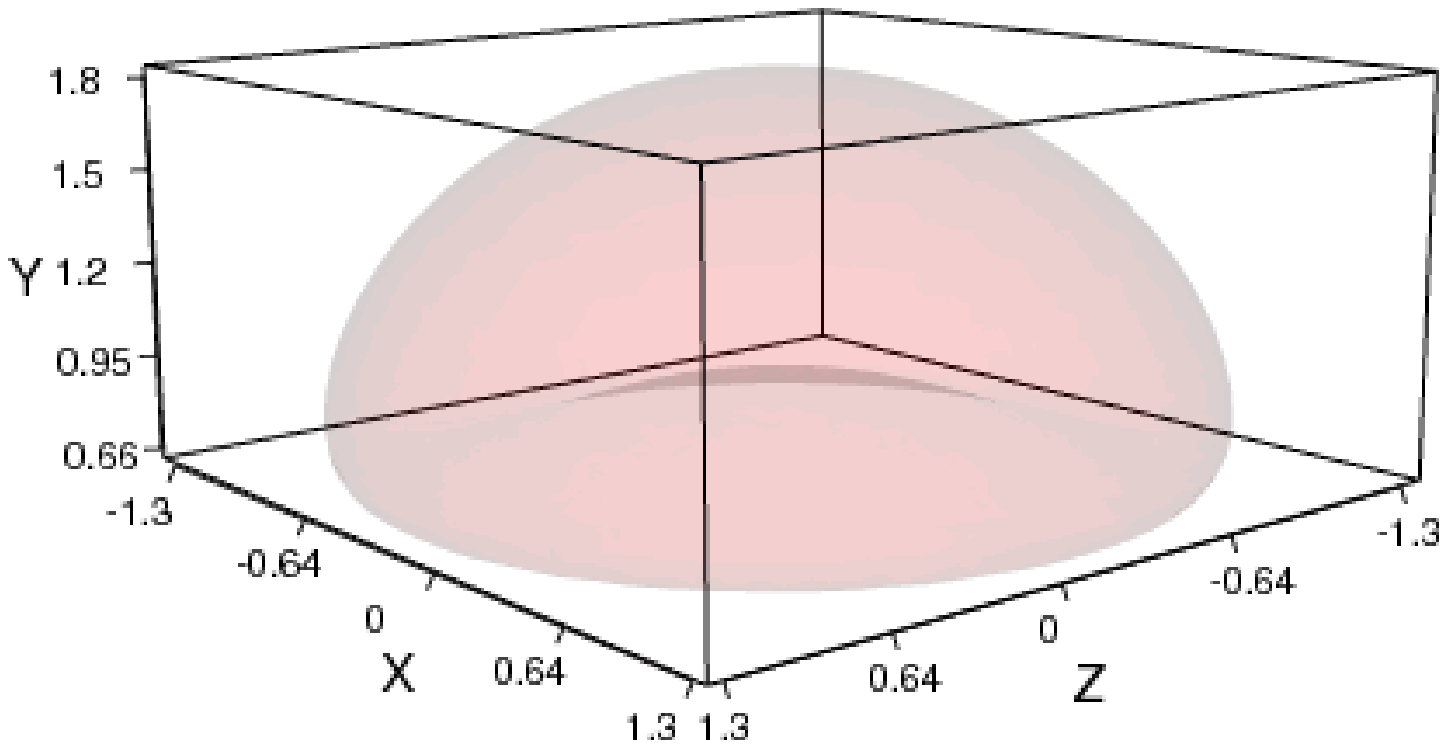}
\caption{Evolution of three energy density levels of the
electromagnetic knot of the Hopf fibration. From left to right and top to bottom,
$T=0$, $T=0.5$, $T=1$, $T=1.5$. Coordinates $(X, Y, Z)$ are dimensionless and are
related to the physical coordinates $(x,y,z)$ through $(X,Y,Z) =
(x,y,z)/L_{0}$. The levels correspond to values $0.1$, $0.2$ and $0.3$ of the 
density energy in $a / (\mu_{0} L_{0}^4)$ units (see Eq. (\ref{11.38})).
The density energy levels displayed are $0.1$, $0.2$ and $0.3$ in increasing order
of colour intensity. At $T=1.5$, the levels $0.2$ and $0.3$ are not present.} 
\label{fig1}
\end{figure}

The energy density of the electromagnetic field (\ref{11.35}) is given by,
\begin{equation}
U ({\bf r},t)=\frac{\varepsilon_{0} E^{2}}{2} + \frac{B^{2}}{2
\mu_{0}}=\frac{a}{4\pi ^{2} \mu_{0} L_{0}^4}
\frac{(1+X^{2}+(Y+T)^{2}+Z^{2})^{2}}{(A^{2}+T^{2})^{3}}.
\label{11.38}
\end{equation}
The maximum of the energy density is located at
$X=Z=0$ during time evolution. The function $U$ is symmetric in the
coordinates $X$ and $Z$. In Figure \ref{fig1}, we show some isosurfaces of
the energy density $U$ for times $T=0, 0.5, 1, 1.5$. The energy density levels 
represented are $0.1$, $0.2$ and $0.3$ in $a/ (\mu_{0} L_{0}^4)$ units. It can 
be seen how the isosurfaces spread as the energy density goes to zero. For time
$T=1.5$, the $0.3$ and $0.2$ levels have dissapeared. Note that the total 
electromagnetic energy of the knot
\begin{equation}
 {\cal E} = \int U \, d^3 r = \frac{2 a}{\mu_{0} L_{0}} , \label{totalenergy}
\end{equation}
remains constant. At $T=0$, the energy
density has spherical symmetry and its maximum is located at the
origin. As time increases, the symmetry is broken and the maximum is 
located at $X=Z=0$ and $Y$ close to $T$. Approximately, the position of the
maximum of the energy density can be found up to $T=1$ at $X=Z=0$, 
$Y=T(1+6 T^2 )/(2+6 T^2 )$.

The energy density of the knot extends to infinity, but we can define a mean 
quadratic radius of the energy distribution as
\begin{equation}
 <r^2 > = \frac{\int ({\bf r} - {\bf r}_{max})^2 U \, d^3 r}{\int U \, d^3 r},
\label{meanradius}
\end{equation}
where ${\bf r}_{max}$ is the position of the maximum of the distribution. At 
$T=0$, the distribution has spherical symmetry and the mean quadratic radius of the 
distribution is given by $\sqrt{<r^2 >} = L_{0}$. The maximum has a value of
the (dimensionless) energy density equal to $16/\pi^2$ and at a distance equal to the 
mean quadratic value, the dimensionless energy density is $1/\pi^2$. This means that, initially, 
more than $70 \%$ of the energy is locallized inside a sphere of radius $L_{0}$ centered at the origin. 
As time evolves the mean quadratic radius of the 
distribution spreads out (its value at $t=L_{0}/c$ or $T=1$ is about $1.1 \, L_{0}$) and the position 
of the center is at $(0,7/8,0)$. Note that the distribution is not well characterized as a sphere as time
evolves.

One can also compute the Poynting vector ${\bf P}$ of the electromagnetic field, obtaining
\begin{equation}
 {\bf P} = \int \frac{{\bf E} \times {\bf B}}{\mu_{0}} \, d^3 r = \left( 0, \frac{a c}{2 \mu_{0} L_{0}}, 0 
\right).
\label{momentum}
\end{equation}
As can be seen, it has a single contribution along the $y$-axis. This explains why the maximum of the energy 
density moves along this axis for the electromagnetic knot studied in this paper.

\section{Relativistic motion of charges in the electromagnetic
knot of the Hopf fibration}

Now we apply the electromagnetic knot studied in the previous
section to the following situation. Suppose that this knot has
been created in certain region of the space, so we have at $t=0$ 
a knot initially centered at the origin, that moves with time 
along the $y$-axis as we discussed in the previous section. Let us 
assume there are free electrons
with negligible initial velocities in units of the light
velocity $c$. We consider the evolution of these electrons under
the electromagnetic knot field. 

The velocity of the electrons increases
by the action of the electromagnetic field, so we will solve the
relativistic equation for the motion of single electrons
\cite{Landau}, considered as test particles which do not affect the value
of the electromagnetic field obtained from the Hopf fibration. The equation 
to be considered is
\begin{equation}
\frac{d {\bf v}}{dt} = - \frac{e}{m} \sqrt{1 - \frac{v^2}{c^2}}
\left( {\bf E} + {\bf v} \times {\bf B} - \frac{1}{c^2 } {\bf v} (
{\bf v} \cdot {\bf E} ) \right), \label{relmotion1}
\end{equation}
where $e = 1.6 \times 10^{-19}$ C is the electron charge, and
$m=9.1 \times 10^{-31}$ kg is its rest mass. Using dimensionless
quantities and the expressions (\ref{11.35}) for the
electromagnetic knot of the Hopf fibration, we find
\begin{eqnarray}
\frac{d {\bf V}}{dT} &=& - \frac{e \sqrt{a}}{\pi m c L_{0}}
\frac{\sqrt{1 - V^2}}{(A^2 +T^2 )^3} \nonumber \\ &.& 
\left( Q{\bf H}_{2} - P{\bf H}_{1} + {\bf V} \times (Q{\bf H}_{1} + P{\bf H}_{2}) 
- {\bf V} ( {\bf V} \cdot (Q{\bf H}_{2} - P{\bf
H}_{1}) \right). \label{relmotion2}
\end{eqnarray}
Different possible physical situations can be studied with
equation (\ref{relmotion2}) by changing the value of the dimensionless prefactor 
$g = e\sqrt{a}/(\pi mc L_{0})$. For electrons, taking into account that the
total energy of the field is ${\cal E} = 2 a /(\mu_{0} L_{0})$ and that the size of
the knot is characterized initially by $L_{0}$, the prefactor in equation 
(\ref{relmotion2}) can be written as
\begin{equation}
g = \frac{e \sqrt{a}}{\pi m c L_{0}} \approx 0.15 \, \sqrt{\frac{{\cal E}}{L_{0}}},
 \label{relmotion3}
\end{equation}
where ${\cal E}$ is measured in Joules and $L_{0}$ in meters.

\begin{figure}
\centering
\includegraphics[width=0.45\textwidth]{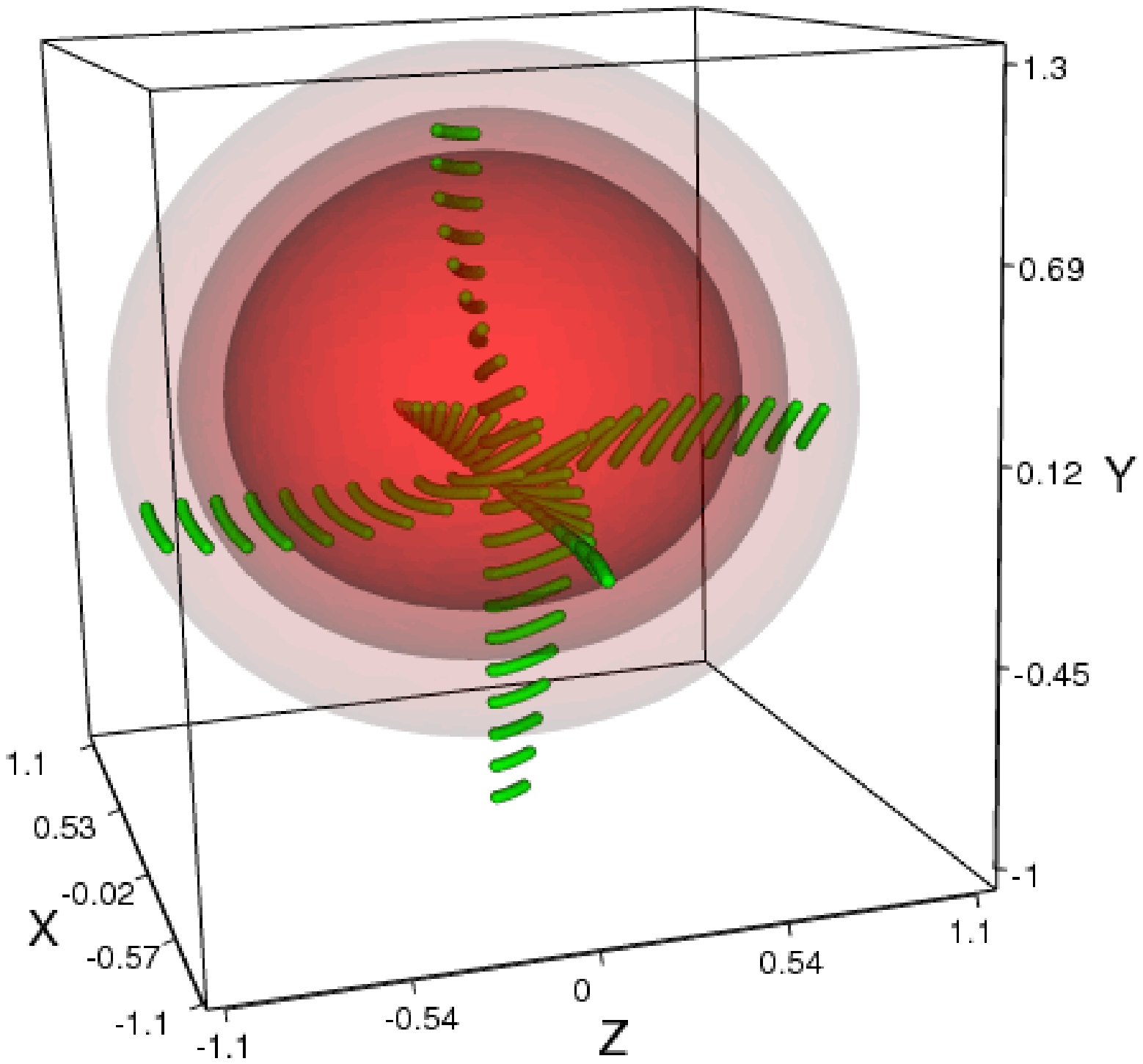}
\includegraphics[width=0.45\textwidth]{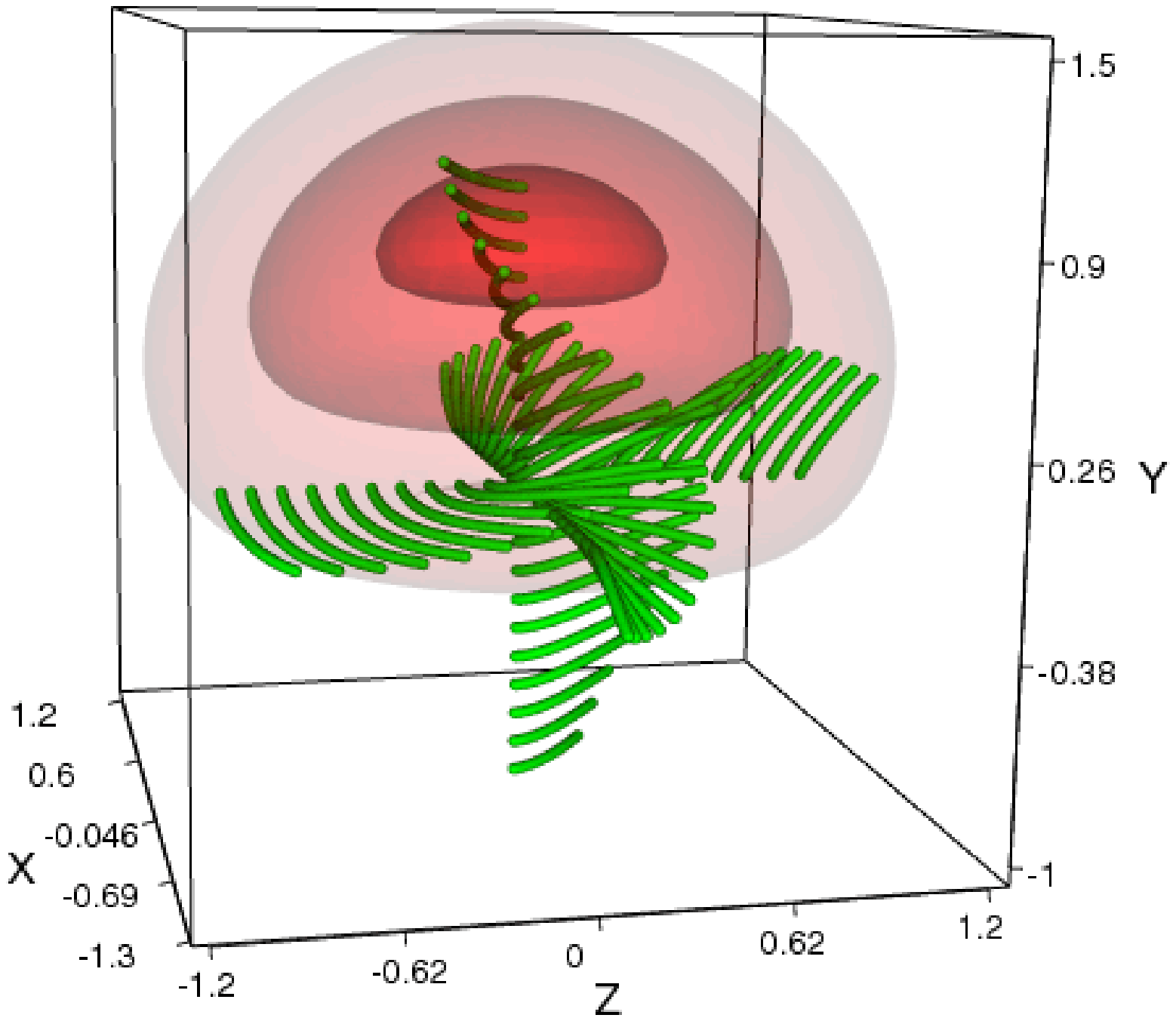}
\includegraphics[width=0.45\textwidth]{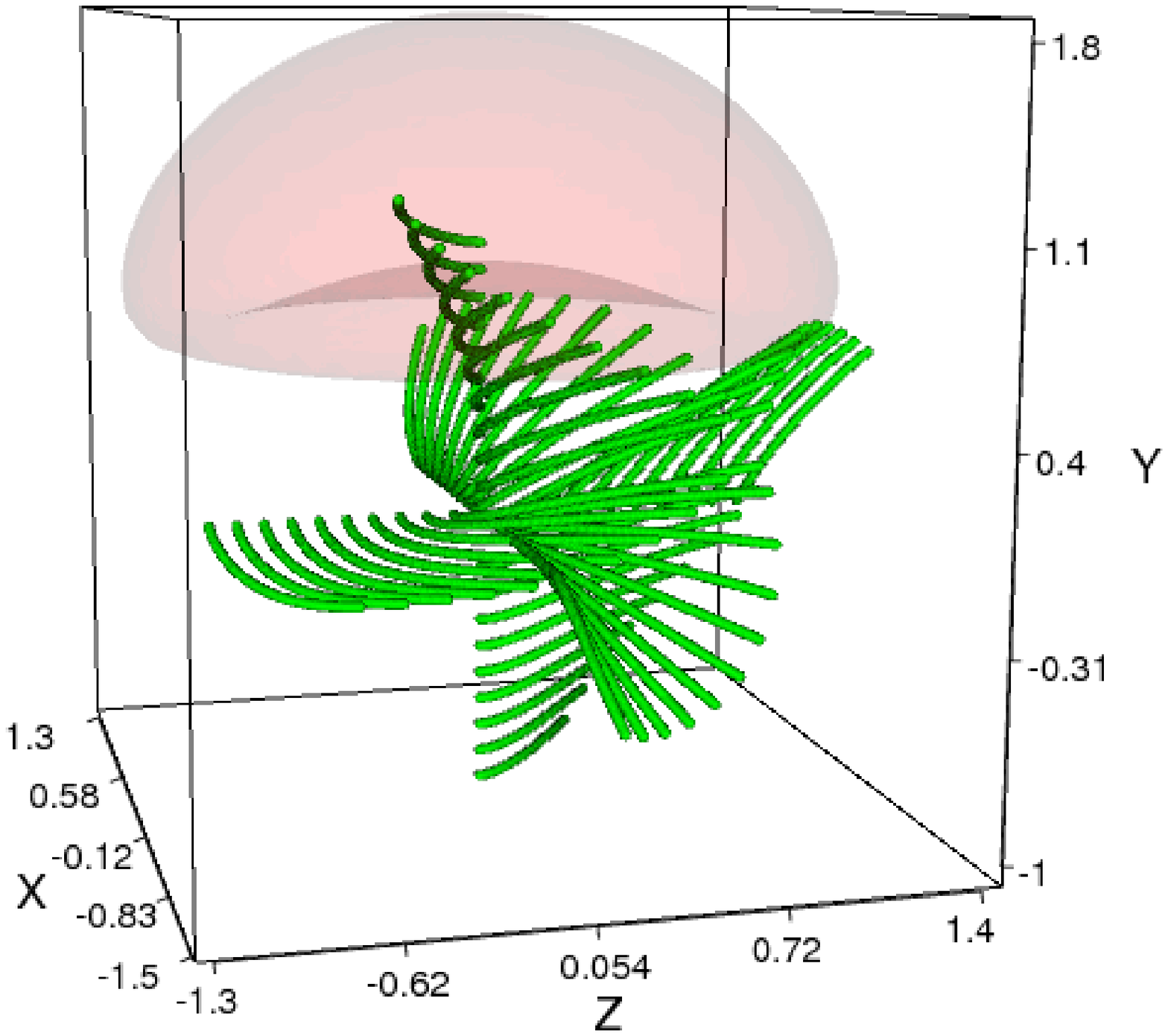}
\caption{Evolution of the position of electrons under the influence of the
electromagnetic knot of the Hopf fibration with $g=1$. From left to right and top to bottom,
$T=0.5$, $T=1$, $T=1.5$. The energy levels shown correspond to values $0.1$, $0.2$ and $0.3$ of the 
density energy in $a / (\mu_{0} L_{0}^4)$ for the same times (see figure \ref{fig1}). The initial velocity 
of the electrons is $V_{0}=0$. The electron initial positions considered are given by 
${\bf R}_{0} = \left( R_{i} , 0 ,0 \right)$, with  $R_{i}= \pm 0.1 , \pm 0.2 , \ldots , \pm 1$, 
by ${\bf R}_{0} = \left( 0 , R_{i} , 0 \right)$, and by ${\bf R}_{0} = \left( 0 , 0, R_{i} \right)$.
After $T=1.5$ these electrons moves almost freely. The final values of the
velocities at $T=1.5$ run from $V_{min} = 0.5300$ to $V_{max} = 0.8410$ in units of the light velocity $c$.} 
\label{fig2}
\end{figure}

In figure \ref{fig2}, the evolution of the position of electrons can be seen when the prefactor
is $g=1$. The initial velocity of the electrons is $V_{0}=0$. We have taken for these electrons
initial positions along the $x$-, $y$- and $z$-axis. In each axis, the dimensionless position is
$\pm 0.1, \pm 0.2, \ldots \pm 1$. In the figure, the electron trajectories are plotted from $T=0$ up to
$T=0.5$ (in the first pannel), up to $T=1$ (in the second pannel), and up to $T=1.5$ (in the third pannel). 
We also include some energy density levels of the electromagnetic knot at the
time considered, corresponding to values $0.1$, $0.2$ and $0.3$ in $a / (\mu_{0} L_{0}^4)$ units (see the 
caption of figure \ref{fig1} for more details). By doing so, in figure \ref{fig2} we see not only the evolution
of the position of some free electrons, but also the region in which the interaction with the electromagnetic 
knot (whose center moves along the $y$-axis with a velocity close to $c$) is more important. This fact explains 
some features of the behaviour of the electron trajectories. At the beginning, up to $T=0.5$, the region in which
the electrons are is influenced by high values of the knot energy density, so that the electrons are accelerated and
their trajectories are curved by the influence of the electromagnetic force. From $T=0.5$ up to $T=1$, the center of 
the knot has moved along the $y$-axis and many electrons are now in a region in which the electromagnetic density is
smaller. However, the electromagnetic force is still important so that many electrons tend to follow the motion of the
center of the knot (a small fraction of the electrons seem to be curved towards the opposite direction). This is 
also the situation from $T=1$ up to $T=1.5$, during which electrons reach high values of the velocity, in the range 
from $V_{min}=0.5300$ to $V_{max}=0.8410$ in units of the light velocity $c$. After this time, the influence of the
electromagnetic knot on the electron trajectories is smaller, and they can be considered as almost free. 

\begin{figure}
\centering
\includegraphics[width=0.45\textwidth]{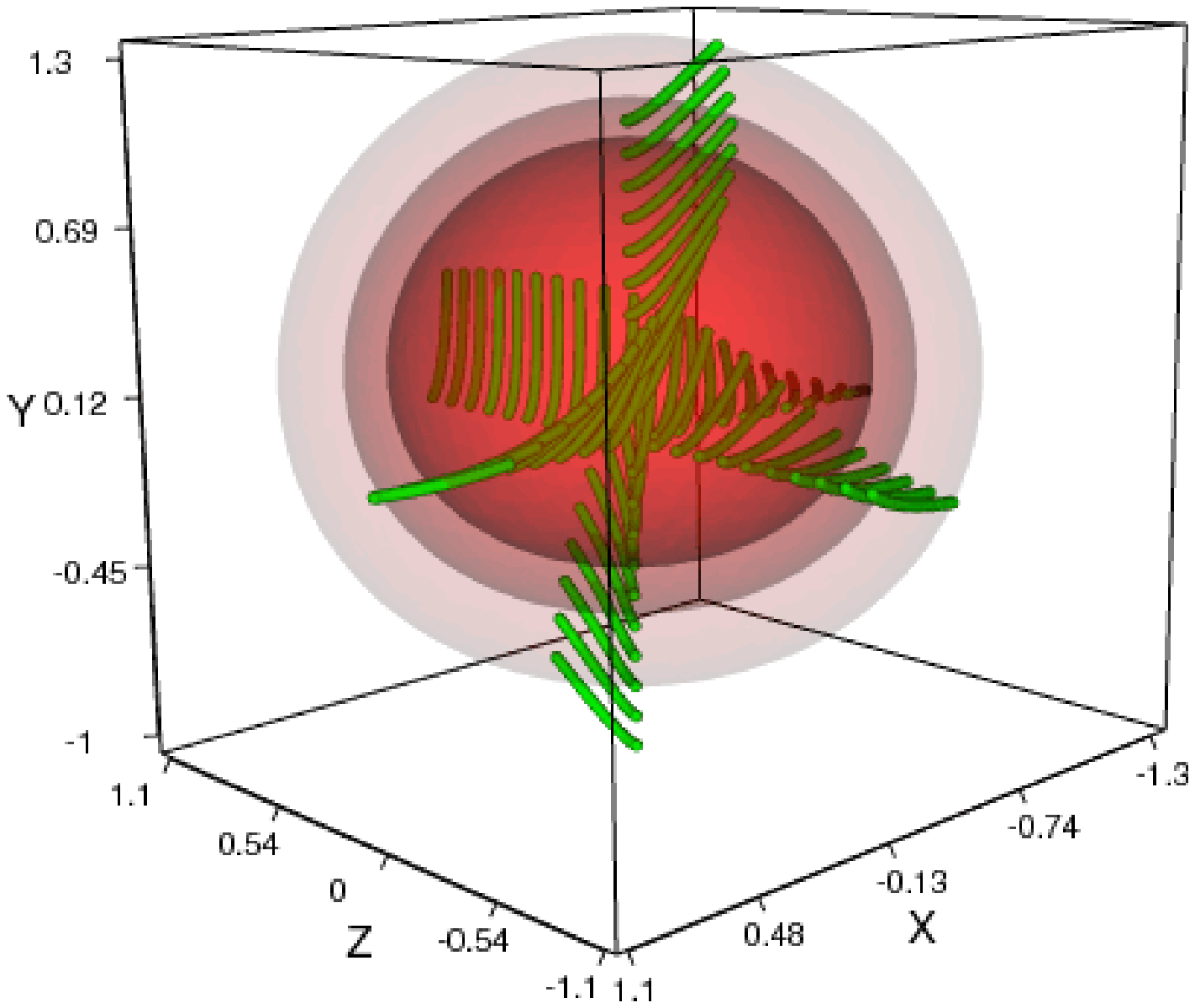}
\includegraphics[width=0.45\textwidth]{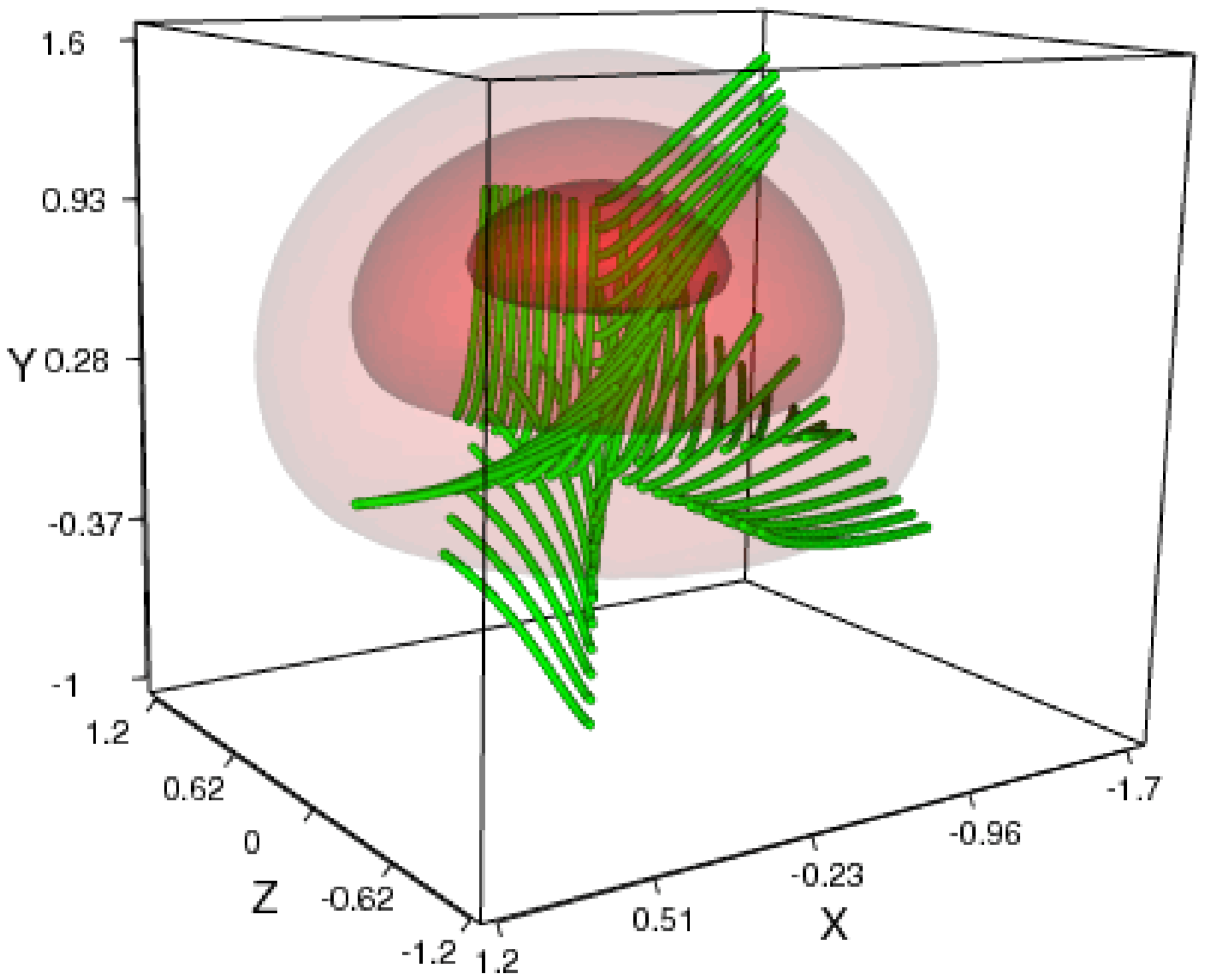}
\includegraphics[width=0.45\textwidth]{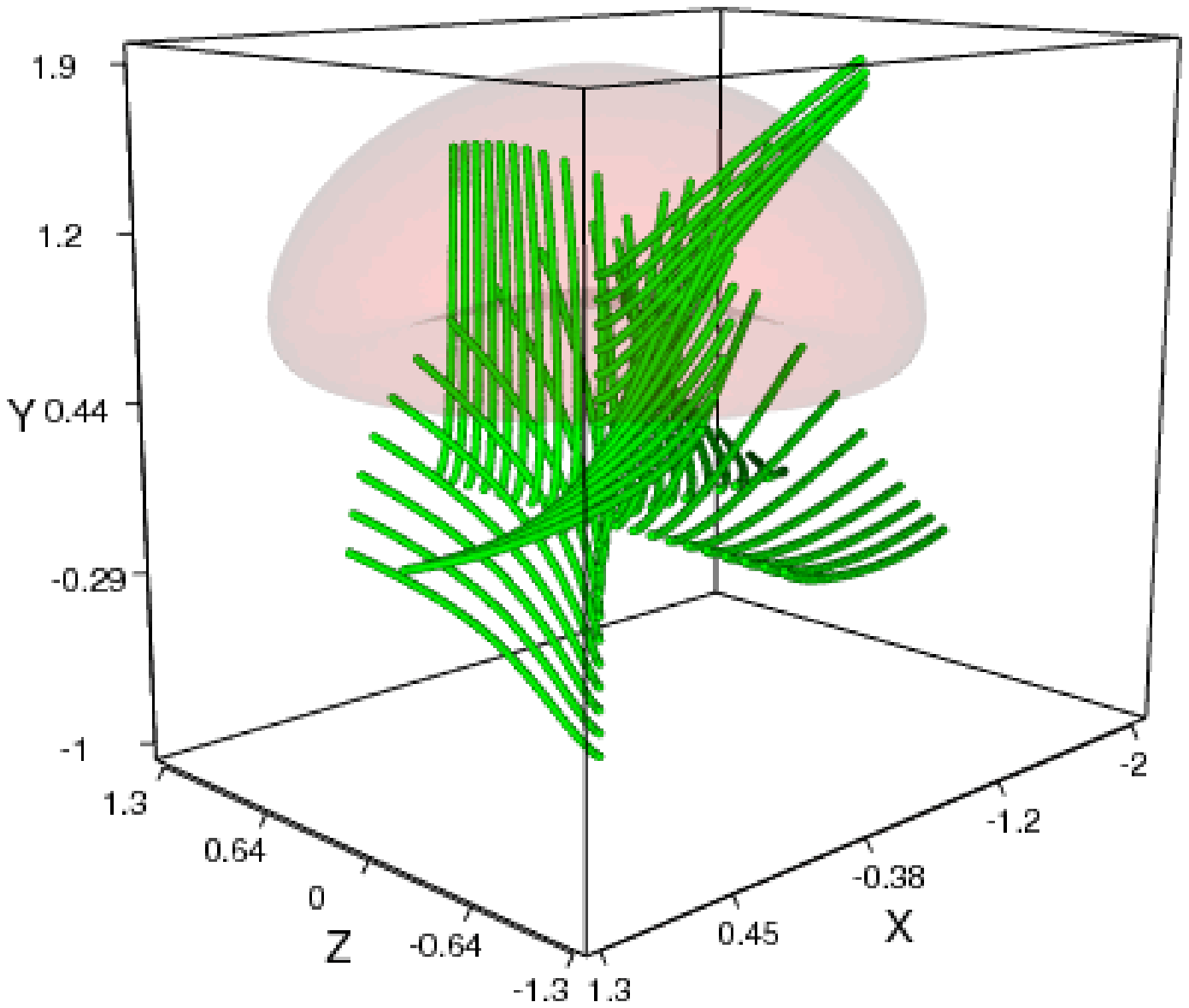}
\caption{Same situation as in figure \ref{fig2} with a value $g=10$. From left to right and top to bottom,
$T=0.5$, $T=1$, $T=1.5$. The energy levels shown correspond to values $0.1$, $0.2$ and $0.3$ of the 
density energy in $a / (\mu_{0} L_{0}^4)$ for the same times (see figure \ref{fig1}). The initial velocity 
of the electrons is $V_{0}=0$. The electron initial positions considered are given by 
${\bf R}_{0} = \left( R_{i} , 0 ,0 \right)$, with  $R_{i}= \pm 0.1 , \pm 0.2 , \ldots , \pm 1$, 
by ${\bf R}_{0} = \left( 0 , R_{i} , 0 \right)$, and by ${\bf R}_{0} = \left( 0 , 0, R_{i} \right)$. The final 
values of the velocities at $T=1.5$ run from $V_{min} = 0.9684$ to $V_{max} = 0.9942$ in units of the light velocity $c$.} 
\label{fig3}
\end{figure}

In figure \ref{fig3}, the evolution of the position of electrons is studied when the prefactor
is $g=10$. The initial velocity of the electrons is $V_{0}=0$, and the initial positions in each axis
are given by $\pm 0.1, \pm 0.2, \ldots \pm 1$ in units of the knot size $L_{0}$ as before. As in figure \ref{fig2}, 
the electron trajectories are plotted in figure \ref{fig3} from $T=0$ up to $T=0.5$ (in the first pannel), 
up to $T=1$ (in the second pannel), and up to $T=1.5$ (in the third pannel), and we include some energy 
density levels of the electromagnetic knot at the time considered, corresponding to values $0.1$, $0.2$ and $0.3$ in $a / (\mu_{0} L_{0}^4)$ units. Up to $T=0.5$, the electrons get higher acceleration and
their trajectories are more curved by the influence of the electromagnetic force than in the case of $g=1$ (figure 
\ref{fig2}). From $T=0.5$ up to $T=1$, all the electrons tend to follow the motion of the
center of the knot, a situation much clearer from $T=1$ up to $T=1.5$. The final velocities of these electrons are
in the range from $V_{min}=0.9684$ to $V_{max}=0.9942$ in units of the light velocity $c$.

\begin{figure}
\centering
\includegraphics[width=0.45\textwidth]{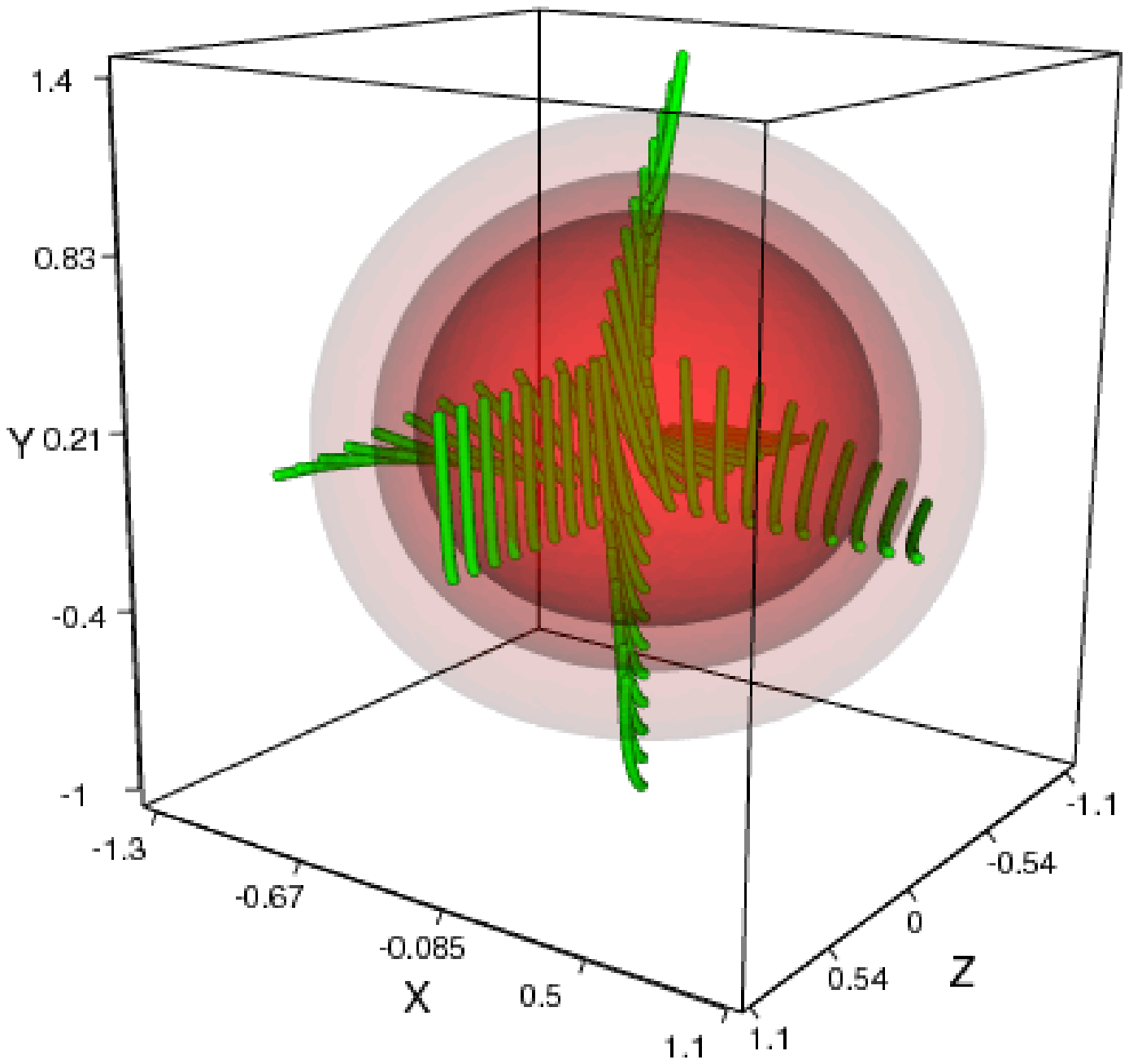}
\includegraphics[width=0.4\textwidth]{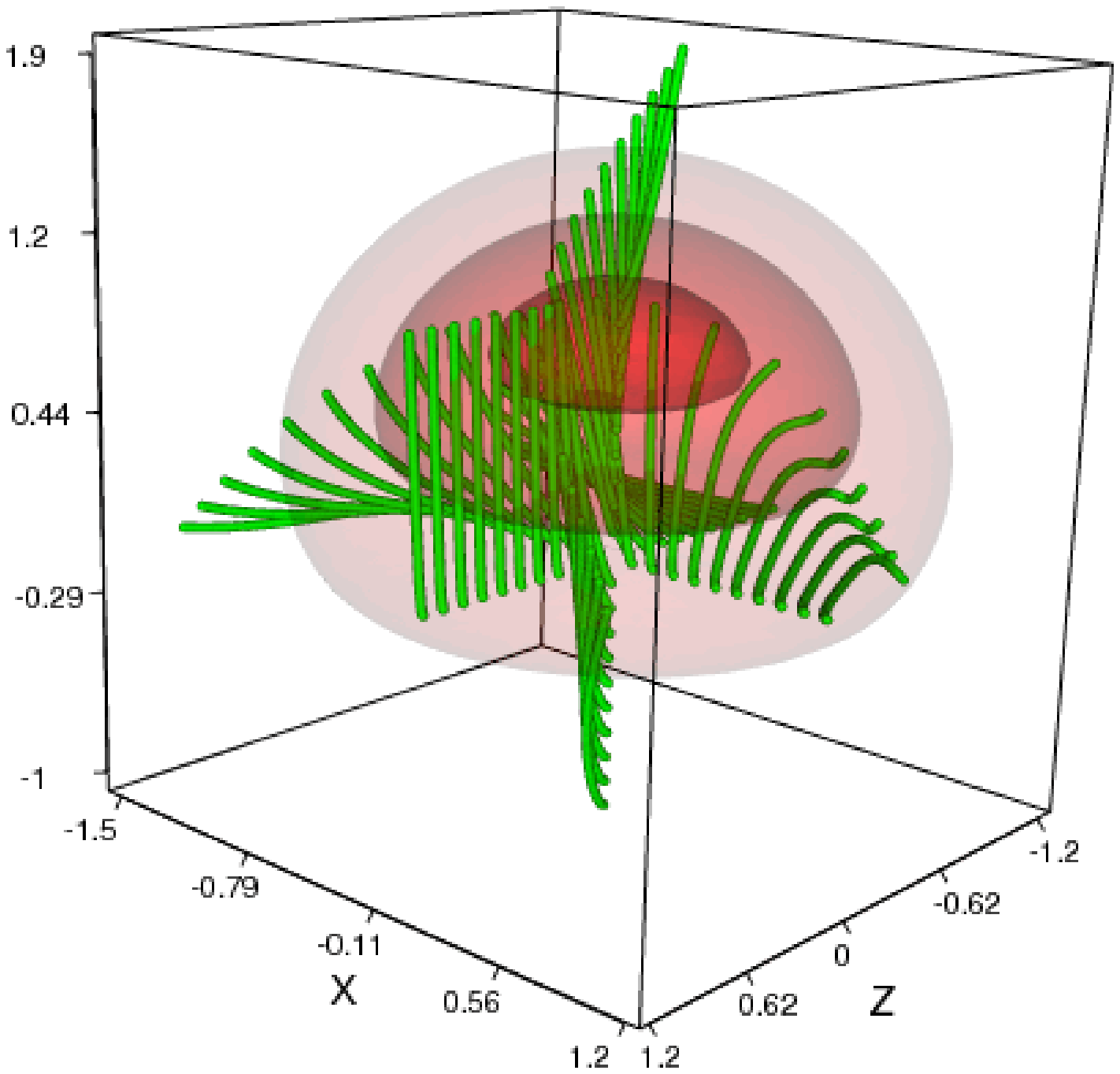}
\includegraphics[width=0.5\textwidth]{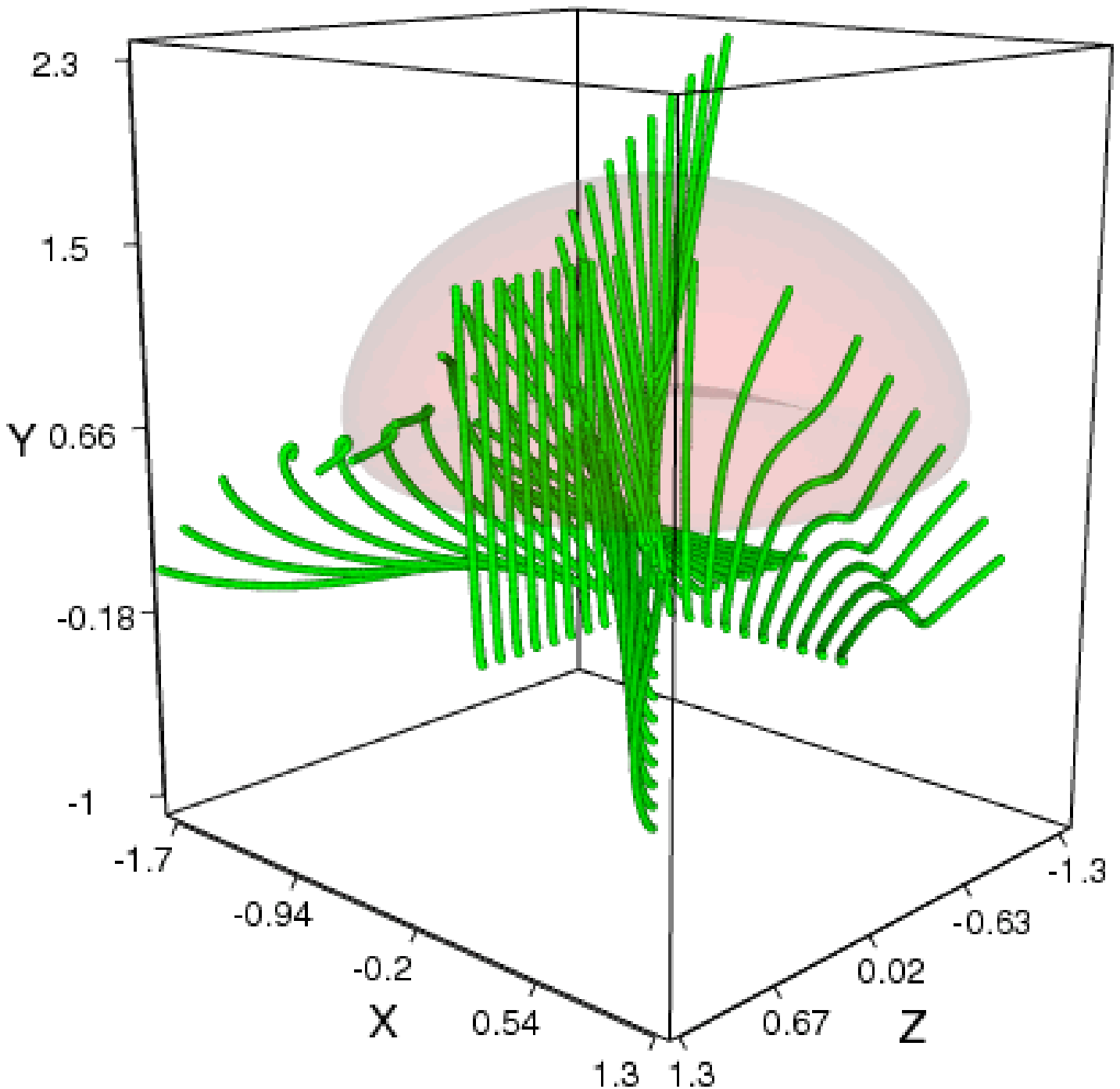}
\caption{Same situation as in figures \ref{fig2} and \ref{fig3} with a value $g=100$. From left to right and
top to bottom, $T=0.5$, $T=1$, $T=1.5$. The energy levels shown correspond to values $0.1$, $0.2$ and $0.3$ of the 
density energy in $a / (\mu_{0} L_{0}^4)$ for the same times. The initial velocity 
of the electrons is $V_{0}=0$. The electron initial positions considered are given by 
${\bf R}_{0} = \left( R_{i} , 0 ,0 \right)$, with  $R_{i}= \pm 0.1 , \pm 0.2 , \ldots , \pm 1$, 
by ${\bf R}_{0} = \left( 0 , R_{i} , 0 \right)$, and by ${\bf R}_{0} = \left( 0 , 0, R_{i} \right)$. The final 
values of the velocities at $T=1.5$ run from $V_{min} = 0.9870$ to $V_{max} = 0.9999$ in units of the light velocity $c$.} 
\label{fig4}
\end{figure}

In figure \ref{fig4}, the evolution of the position of electrons can be seen when the prefactor
is $g=100$. The initial velocity and initial position of the electrons are the same as in the cases $g=1$ (figure \ref{fig2})
and $g=10$ (figure \ref{fig3}). As in previous cases, electron trajectories are plotted in figure \ref{fig4} from $T=0$ up to 
$T=0.5$ (in the left-hand side pannel), up to $T=1$ (in the center pannel), and up to $T=1.5$ (in the right-hand side 
pannel), and we include some energy density levels of the electromagnetic knot at the time considered, corresponding to 
values $0.1$, $0.2$ and $0.3$ in $a / (\mu_{0} L_{0}^4)$ units. The global behaviour of the electron trajectories is 
very similar to the case in which $g=10$, but now the electrons are even more accelerated and their trajectories are even 
more curved by the influence of the electromagnetic force. All the electrons clearly tend to follow the motion of the 
center of the knot along the $y$-axis, with final velocities in the range from $V_{min}=0.9870$ to $V_{max}=0.9999$ in units 
of the light velocity $c$.

\section{Conclusions}

In this paper we have considered the classical relativistic motion of charged particles 
in a knotted electromagnetic field. We have seen how to construct electromagnetic knots
from maps between the three-sphere and the two-sphere. In the particular case of the Hopf map,
whose fibres are mutually linked, we have written the expression for the electromagnetic field. 
This is a solution of the Maxwell equations in vacuum such that any pair of electric lines is a
link and any pair of magnetic lines is a link. We have considered some properties of this field,
in particular the electromagnetic energy density. We have seen that the most part of the energy 
density of the knot is, at $t=0$, localized into a sphere whose radius is the mean quadratic 
radius of the energy density distribution. As time evolves, the spherical symmetry is broken and 
the mean quadratic radius of the distributios spreads out.

We have considered the relativistic motion of electrons (considered as point particles) in the 
electromagnetic field of the Hopf map. We have computed the trajectories of the electrons starting 
with zero initial velocities, and
we have seen that these electrons are strongly accelerated by the electromagnetic force, 
becoming ultrarelativistic in a period of time that depends on the knot size.

Finally we consider that a deeper understanding of the interaction between electromagnetic knots 
and test particles could be useful to design experiments to produce knots in the laboratory.  

\section*{Acknowledgements}

The authors thank support from the Spanish Ministerio de
Educaci\'on y Ciencia under project AYA2009-14027-C07-04.


\vspace{2pc}

\begin{thebibliography}{123456}

\bibitem{Irv08} W. T. M. Irvine and D. Bouwmeester, {\it Nature Phys.} {\bf 4}, 716-720 (2008).

\bibitem{Ran89} A. F. Ra\~nada, {\it Lett. Math. Phys.} {\bf 18}, 97-106 (1989).

\bibitem{Ran92} A. F. Ra\~nada, {\it J. Phys. A: Math. Gen.} {\bf 25}, 1621-1641 (1992).

\bibitem{Ran95} A. F. Ra\~nada and J. L. Trueba, {\it Phys. Lett. A} {\bf 202}, 337-342
(1995).

\bibitem{Ran97} A. F. Ra\~nada and J. L. Trueba, {\it Phys. Lett. A} {\bf 235}, 25-33
(1997).

\bibitem{Ran98} A. F. Ra\~nada and J. L. Trueba, {\it Phys. Lett. B} {\bf 422}, 196-200 (1998).

\bibitem{Ran03} A. F. Ra\~nada, {\it Phys. Lett. A} {\bf 310}, 134-144 (2003).

\bibitem{Ran06} A. F. Ra\~nada and J. L. Trueba, {\it Found. Phys.} {\bf 36}, 427-436 (2006).

\bibitem{Hopf} H. Hopf, {\it Math. Ann.} {\bf 104}, 637–665 (1931).

\bibitem{Ran01} A. F. Ra\~nada and J. L. Trueba, in: {\it Modern
Nonlinear Optics. Part III}, ed. M. Evans (Wiley \& Sons, New
York, 2001), pp. 197-254.

\bibitem{Tru09} J. L. Trueba and M. Array\'as, {\it J. Phys. A: Math. Theor.} {\bf 42}, 282001 (2009).
 
\bibitem{Landau} L. D. Landau and E. M. Lifshitz, {\it The Classical
Theory of Fields} 4th Ed. (Elsevier, Amsterdam, 1975).

\end{thebibliography}
\end{document}